\documentclass[twocolumn,superscriptaddress,aps,prl]{revtex4}
\usepackage{graphicx}
\usepackage{color}
\usepackage[normalem]{ulem}

\begin{document}

\title{Phase diagram of the ground states of DNA condensates}

\author{Trinh X. Hoang}
\affiliation{Institute of Physics, Vietnam
Academy of Science and Technology, 10 Dao Tan, Ba Dinh, Hanoi, Vietnam}

\author{Hoa Lan Trinh}
\affiliation{Institute of Physics, Vietnam
Academy of Science and Technology, 10 Dao Tan, Ba Dinh, Hanoi, Vietnam}

\author{Achille Giacometti}
\affiliation{Dipartimento di Scienze Molecolari e Nanosistemi, Universita' Ca'
Foscari Venezia, I-30123 Venezia, Italy}

\author{Rudolf Podgornik}
\affiliation{
Department of Theoretical Physics, J. Stefan Institute and
Department of Physics, Faculty of Mathematics and Physics, University of
Ljubljana - SI-1000 Ljubljana, Slovenia, EU}
\affiliation{Department of Physics, University of Massachusetts, Amherst, MA 01003, USA}

\author{Jayanth R. Banavar}
\affiliation{Department of Physics, University of Maryland, College Park, 
Maryland 20742, USA}

\author{Amos Maritan}
\affiliation{Dipartimento di Fisica e Astronomia, Universit\`a di Padova, CNISM and INFN, via Marzolo 8,
I-35131 Padova, Italy}

\begin{abstract}
Phase diagram of the ground states of DNA in a bad solvent is studied for a
semi-flexible polymer model with a generalized local elastic bending potential
characterized by a nonlinearity parameter $x$ and effective self-attraction
promoting compaction. $x=1$ corresponds to the worm-like chain model.
Surprisingly, the phase diagram as well as the transition lines between the
ground states are found to be a function of $x$. The model provides a simple
explanation for the results of prior experimental and computational studies
and makes predictions for the specific geometries of the ground states.
The results underscore the impact of the form of the microscopic bending 
energy at macroscopic observable scales.
\end{abstract}

\maketitle

The non-linear elasticity of DNA at short length scales, probed by
non-equilibrium DNA cyclization experiments \cite{Cloutier,Vologotskii}, AFM
imaging on surfaces \cite{Wiggins}, as well as by equilibrium mechanically
constrained DNA experiments \cite{Zocchi1,Zocchi2,Zocchi3}, is still not well
understood. While the conclusions regarding the first two methods for
probing non-linear ds-DNA elasticity have been criticized
\cite{Vologodski3}, the experiments on stressed DNA ring molecules
are performed by a different methodology based on thermodynamic methods {\sl
via} DNA high-curvature states through partial hybridization of a ss-DNA loop
with a linear complementary strand \cite{Zocchi3}. This methodology does not
depend on thermal fluctuations to realize high-curvature states and thus
appears to have a far better accuracy and reliability. The non-linearity in
this latter case is clearly present and is captured in a single parameter
describing the onset of DNA kink formation. This clearly exhibited elastic
non-linearity is taken as the motivation for the present study that attempts to
derive the macroscopic consequences of this interesting microscopic elastic
behavior of DNA.

Another interesting facet of DNA behavior is that under specific solution conditions, it condenses into highly compact
structures with pronounced symmetry \cite{Gosule,Bloomfield,Hud05,RP1}. This
condensation phenomenon serves as an example of high polymer density packing in
biology and of polymer phase transitions and phase separations in general,
being relevant also for artificial gene delivery \cite{Livolant,Nancy}.
Several distinct morphologies of the DNA condensates have been observed
including a toroid, a spheroid as well as a rod-like configuration
\cite{Hud05,Arscott,Golan,Pinto}.  Our principal goal is to explore the rich
interplay between the strong tendency for compactness, arising from the
presence of multivalent cations or osmolytes in the solution, and the intrinsic
stiffness of the DNA molecule promoting the chain to be locally straight. 
In the absence of stiffness, one would expect the chain to adopt a densely
compact spheroidal globule configuration.  The key issue is to understand how
local stiffness and the detailed way it enters the elastic energy result in
the spheroidal configuration becoming unstable w.r.t. other lower energy
configurations. In particular, one would like to map out a phase diagram and
understand the different condensate geometries. 

The topology of a toroid can be modified in at least two distinct ways. 
One is to cut and elongate it making it rod-like and the other
is by filling up the hole. Toroids of size larger than $\sim$200
nm have been observed. In a detailed study \cite{Conwell}, the toroid mean
diameter was found to vary between 30 and 100 nm depending on the solution
conditions, while the toroid thickness was in the range from 10 to 70 nm. The
geometry of a toroid can be characterized by the ratio between its thickness
and its diameter. This ratio was typically found to vary for given solution
conditions but had a maximum value in a small interval between 0.7 and 0.9.
Ratios close to 1 have been observed for very large toroids \cite{Conwell2004}.
A ratio of 1 corresponds to a toroid with no hole.  This is the same
topology as a sphere but has different geometry than an isotropic spheroid.

Our study here essentially focuses on the relative stability of different
ground states, disregarding the elastic fluctuations around these ground
states. An important lesson learned in the study of critical phenomena is that
details at the microscopic level often do not matter for the behavior near a
critical point.  Strikingly, however, we find that simple modifications
in the form of the elastic energy penalty at the microscopic scale have
considerable macroscopic consequences. The nature of the phase diagram as well
as of the phase transitions is found to depend on the details of the
local stiffness energy, allowing us to provide a simple explanation for the
geometries of the commonly observed toroidal structures of condensed DNA. 

The simplest and most commonly used model for describing the stiffness of a
chain molecule is the worm-like chain (WLC) model. Consider 
three (1,2,3) consecutive beads along the chain. Let $\theta$ represent the
bending angle, {\sl i.e.}, the angle between vectors 1-2
and 2-3. $\theta$ would be zero for a straight segment. The elastic
(free) energy penalty on the WLC level is then:
\begin{equation}
\label{eq:gWLC}
u = \kappa (1-\cos \theta)^x ,
\end{equation}
where $\kappa$ is the stiffness coefficient per bead and $x=1$.  A recent study
of the ground states of a chain molecule \cite{HoangJCP} using the WLC together
with different types of interactions promoting compaction resulted in a phase
diagram with toroidal and rod-like ground states. Furthermore, the combination
of analytical calculations and computer simulations showed that the shape of
the phase diagram was quite insensitive to the microscopic details. The
challenge is then to construct the simplest model, able to qualitatively
reproduce the key experimental observations, while at the same time also
allowing some flexibility in the predicted behavior, observed in
experiments. 

A value of $x$ larger than 1 corresponds to a softer potential promoting easy
bending. Conversely, the system is stiffer when $x$ is less than 1. Generally,
$x$ cannot be less than 1/2 because of the behavior of the elastic energy for
small $\theta$.  The coupling between different forms of
attractive interaction and thermal fluctuations in WLC polymer collapse was
addressed by Hansen et al. \cite{Hansen}. Schnurr et al. \cite{Schnurr} 
employed Brownian dynamics simulations complemented by analytical theory to
study the dynamical intermediates for a limiting case of WLC model. Lappala and
Terentjev \cite{Lappala} carried out dynamical computational studies of the
compaction of a long chain and observed multiple -- some metastable --
configurations.  Seaton et al. \cite{Landau} studied structural phases
of semi-flexible polymers as a function of temperature/stiffness. The
stabilities of toroidal and rod-like condensates under stretching forces have
been assessed \cite{Cortini}.  Interestingly, DNA-packaging simulations
carried out for a $x=1$ chain yield a stable torus-like structure 
\cite{Marenduzzo}. A careful study by Sakaue and Yoshikawa of a $x=2$
chain dynamics undergoing compaction showed a stable toroidal
phase and a metastable rod-like phase \cite{Sakaue02}.

\begin{figure}
\begin{center}
\includegraphics[width=3.0in]{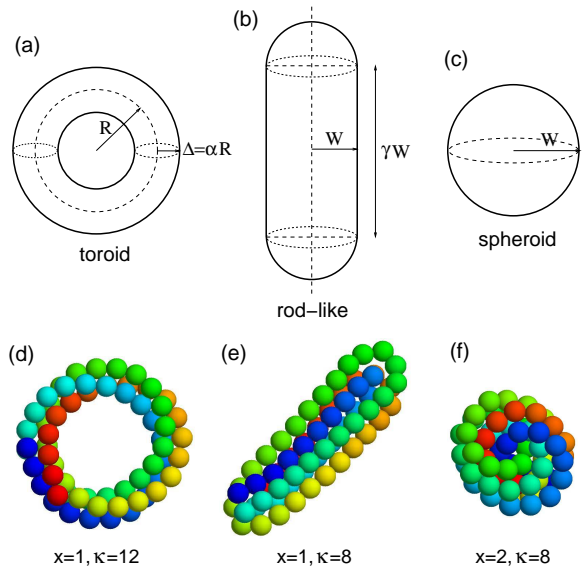}
\end{center}
\caption{\label{fig:model} Schematics of toroidal (a), rod-like (b)
and spheroidal (c) models of DNA condensates.  Ground state conformations
obtained in Monte Carlo simulations of a $N=64$ bead-and-spring model with
$x=1$ generalized elastic potential at moderate (d) and low (e) stiffness and
$x=2$ at low stiffness (f). The conformation (f) is a torus with no hole.}
\end{figure}

Our analysis of the generalized WLC model, as described by Eq. (\ref{eq:gWLC}),
under compaction shows a rich variety of ground states as a function of the
stiffness parameter $x$: a collapsed spheroid, a rod-like configuration that we
model as a cylinder with spherical end caps, a toroid with varying shape
ratio, and a swollen extended phase.  Surprisingly, the transition lines
between these states are found to be characterized by exponents which depend on
$x$. In addition, our analysis predicts the geometry of the stable toroids and
shows a striking effect of the details of the microscopic bending energy
penalty on the macroscopic behavior, relevant for both experiments and
simulations. 

We model DNA as a semi-flexible polymer of length $L=Nb$, formed of $N$
spherical beads of diameter $b$.  Assume that the morphology of DNA condensate
can be either toroidal (Fig. \ref{fig:model}a) or rod-like (Fig.
\ref{fig:model}b).  The toroidal structure is parametrized by its mean radius,
$R$, and the
thickness radius, $\Delta=\alpha R$, with $0<\alpha< 1$. The rod-like structure
is postulated to have a shape of a sphero-cylinder of radius $W$ and the
cylinder length equal to $\gamma W$ with $\gamma \ge 0$. When $\gamma = 0$, one
obtains the spheroidal configuration (Fig. \ref{fig:model}c) as a limiting case
of the rod-like structure. We will show that the spheroidal configuration will
appear as the ground state not only in the limit of $\kappa \rightarrow 0$ or
$L \rightarrow \infty$ for $x<3/2$, but also as a true ground state when
$x>3/2$.

Assume a close-packed hexagonal chain packing in the condensate \cite{Hud05} 
with DNA-DNA interaxial spacing $d$.  The DNA volume packing fraction is 
$\eta=(\pi/\sqrt{12}) (b/d)^2$.  The toroidal mean radius can be expressed 
as $R = \frac{1}{2}\left({Lb^2}/{\pi \eta}\right)^{1/3} \alpha^{-2/3}$,
while the thickness radius of the rod-like condensate is 
$W = \left({3Lb^2}/{4\eta} \right)^{1/3} (4+3\gamma)^{-1/3}$.
For the toroidal condensate, we assume that the chain has a constant radius of
curvature \cite{Hud95} equal to its mean radius $R$. 
The alternative ``spool'' model \cite{spool} for chain wrapping does not change
the characteristics of the phase diagram and yields a higher bending energy
than the constant curvature model.  For the rod-like condensate, we assume that
the radius of curvature is uniformly equal to $W/2$ in the spherical caps of
the sphero-cylinder, whereas the curvature is zero in the cylinder body.

\begin{figure}
\centerline{\includegraphics[width=3.0in]{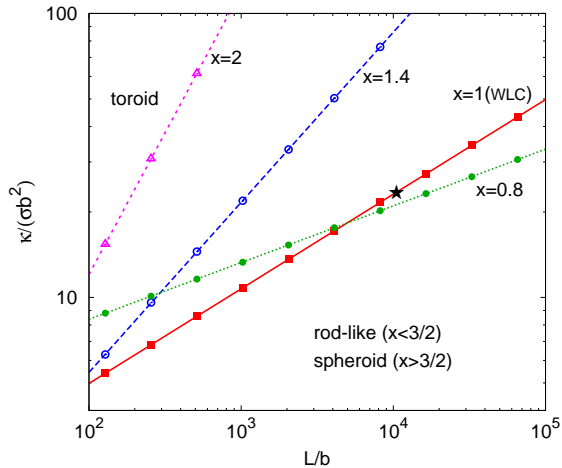}}
\caption{\label{fig:phase}
Ground state phase diagrams in the length-stiffness plane of a semi-flexible
chain with the generalized elastic potential for several values of $x$ as
indicated.  The lines correspond to transitions from a toroidal to a rod-like
condensate for $x<3/2$, and from a toroidal to a spheroidal configuration for
$x>3/2$, being fits with slopes equal to $\frac{2x-1}{3}$ to numerical data
points.  An experimental data point (star) shown on the WLC model transition
line corresponds to the average DNA length of $\sim$63.6 kbp in 90 nm sized
condensates found for a plasmid DNA condensation induced by cobalt(III)
hexammine (CoHex$^{3+}$) with a coexistence of toroid phase and rod-like phase
\cite{Arscott}.
Projection from this point on the vertical axis yields $\sigma \approx 0.266 \,
k_B T$(nm)$^{-2}$, or equivalently, an intermolecular contact interaction
energy of $\varepsilon \approx -\sigma d \approx -0.246$ $k_B T$/bp, by using
the DNA persistence length $ \beta \kappa b = 50$ nm, diameter $b=2$ nm, and
the interaxial spacing $d=2.8$ nm.  The energy $\varepsilon$ roughly agrees
with that for the CoHex$^{3+}$ mediated DNA-DNA interaction obtained by osmotic
stress method ($- 0.21 \pm 0.02 \ k_B T$/bp) \cite{Todd}.
}
\end{figure}

The chain compaction is induced by introducing a surface term in the total
energy of the condensate, with $\sigma > 0$ a surface energy per unit area. For
the toroidal condensate, the total energy including the bending energy is 
then:
\begin{equation}
E_\mathrm{tor} = \kappa \frac{L}{b} \left(\frac{b^2}{2R^2}\right)^x
+ \sigma 4\pi^2 \alpha R^2 ,
\end{equation}
and for the rod-like condensate, respectively:
\begin{equation}
E_\mathrm{rod} = \frac{2^{4+x}}{3} \kappa \eta \left(\frac{b}{W}\right)^{2x-3} 
+ \sigma 2\pi (2 + \gamma) W^2 .
\end{equation}
For a given $L$ and $\kappa$, the minimum energy $E_\mathrm{tor}^*$ and the
optimal ratio $\alpha^*$  for the toroidal condensate are obtained by
minimizing $E$ with respect to $\alpha$. The rod-like condensate can be a
ground state only if $x<\frac{3}{2}$, when $\partial E_\mathrm{rod}/
\partial \gamma=0$ has a positive root, $\gamma^* > 0$.  For $x > \frac{3}{2}$,
there are no positive roots, and the minimum energy is the spheroidal
configuration ($\gamma^*=0$).

By comparing $E_\mathrm{tor}^*$ with $E_\mathrm{rod}^*$, we constructed the
phase diagrams of the ground states in the $\kappa$-$L$ plane (Fig.
\ref{fig:phase}). For a given $L$, one observes a transition from rod-like or
spheroidal to the
toroidal condensate on increasing $\kappa$. For very large stiffness, there is
another transition from the toroidal to an open straight conformation (not
shown in Fig. \ref{fig:phase}). It can be shown that the ratio
$E^*_\mathrm{rod}/E^*_\mathrm{tor}$ is a function of $\sigma/\kappa
L^{(2x-1)/3}$, so the only way to make it equal to 1 while
changing $L$ and $\kappa$ is to have $\sigma/\kappa L^{(2x-1)/3} = $ const.
Thus, at the transition line between the toroidal and the rod-like
condensate, one has:
\begin{equation}
\frac{\kappa}{\sigma} \sim L^\frac{2x-1}{3}.
\label{eq:trans1}
\end{equation}
For $x=1$, one recovers the power law of $L^{1/3}$ \cite{HoangJCP}. 
For $x=0.5$, the transition line would become independent of $L$, but
for this limiting case, both the toroid phase and the
rod-like phase disappear (see below).

Coexisting toroidal and rod-like condensates of similar volume and dimensions
have been observed in experiments \cite{Arscott,Golan} suggesting that those
condensates were close to a phase transition between the two kinds of
structures.  Fig. \ref{fig:phase} shows that fitting the experimental data to
the phase diagram yields a reasonable energy for DNA-DNA interaction.  The
relatively larger volume seen for spheroidal condensates w.r.t.
toroidal ones, as observed in T4 DNA condensation using protamine sulfate as
the condensing agent \cite{Pinto}, agrees with our phase diagram. The
observation of the spheroidal condensates also suggests that the highly charged
protamines, when bound to DNA major grooves \cite{Balhorn}, strongly modify the
DNA elasticity making it incompatible with the WLC model.  Our generalized
elasticity model is applicable also to multimolecular condensates, since due to
a strong sticky hydrophobic interaction between the exposed ends of DNA
fragments, there is no significant free energy contribution from linear
aggregation of the molecules \cite{Zanchetta}.

\begin{figure}
\begin{center}
\includegraphics[width=2.8in]{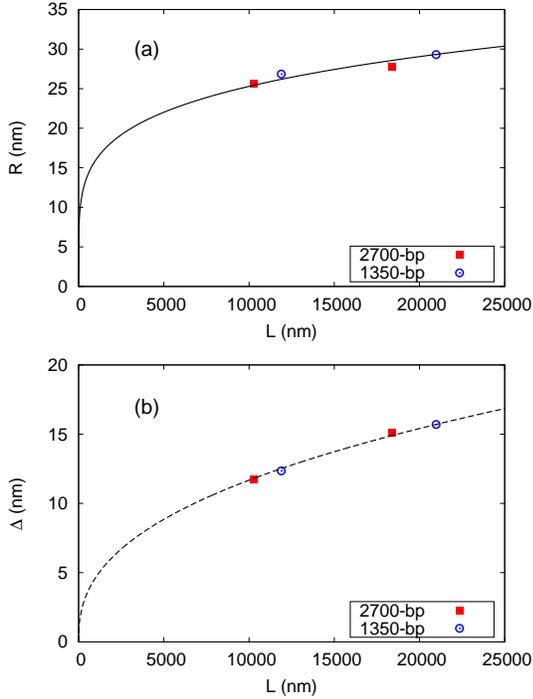}
\end{center}
\caption{\label{fig:size} 
Dependence of the toroidal mean radius $R$ and the toroidal thickness radius
$\Delta$ on the total contour length $L$ of DNA in toroidal condensates.
Experimental data points are fitted to the theoretical results
for the WLC model ($x=1$) as $R \sim L^{1/5}$ (solid line)  and $\Delta \sim
L^{2/5}$ (dashed line) in (a) and (b), respectively. Experimental data 
are taken from Ref. \cite{Arscott} 
for condensation of plasmid DNA fragments of
lengths 2700 bp (squares) and 1350 bp (circles) with [DNA]=10 $\mu$g/mL. The
condensation was induced by 150 $\mu M$ CoHex$^{3+}$ 
and the sizes of the condensates were measured at 2 hours and at 24 hours after 
polyvalent salt addition. The volume of an average toroid was found to increase
by $\sim$ 50\%, between 2 and 24 h. 
$R$ and $\Delta$ are calculated 
based on the measured average toroid outer radius
$R_1$ and inner radius $R_2$ given in the captions of Figs. 3
and 4 of Ref. \cite{Arscott} 
as $R=(R_1+R_2)/2$ and $\Delta=(R_1 - R_2)/2$,
respectively. 
The DNA contour length in a condensate is determined as
$L=8\pi\eta R (\Delta/b)^2$
by assuming a hexagonal close packing of DNA in the condensate
with the volume fraction $\eta=(\pi/\sqrt{12})(b/d)^2$,
where $d=2.8$ nm is the DNA-DNA interaxial spacing,
and $b=2$ nm is the diameter of DNA. 
}
\end{figure}

Interestingly, the geometrical parameters $\alpha^*$ and $\gamma^*$ are also
found to be functions of $\sigma/\kappa L^{(2x-1)/3}$, and thus are constant
along the transition line for a given $x$.  For both types of condensates, the
size can be normalized by $L_c$, the length of the polymer at the transition
line.  For toroidal condensates it then follows
$\alpha^* \sim (L/L_c)^\frac{2x-1}{4x+1}$ 
when $\alpha^* < 1$, whereas for the rod-like condensates, the dependence of
$\gamma^*$ on $L$ is not trivial. For large length ($L \gg L_c$), we remain
with $\gamma^* \sim (L/L_c)^\frac{1-2x}{3}$.
It is also straightforward to show that the toroid mean radius $R$ depends on
chain length as $R \sim L^{\frac{1}{4x+1}}$, and the toroid's thickness scales
with its mean radius as $\Delta \sim R^{2x}$. For $x=1$ one finds that $\Delta
\sim R^2$, meaning that a big toroid is also much thicker than a small
toroid, as commonly observed in experiments.  
Fig. \ref{fig:size} shows that the scaling of $R$ and $\Delta$ with $L$
also agree with available experimental data for condensation of
plasmid DNA \cite{Arscott}.

\begin{figure}
\center
\includegraphics[width=3.0in]{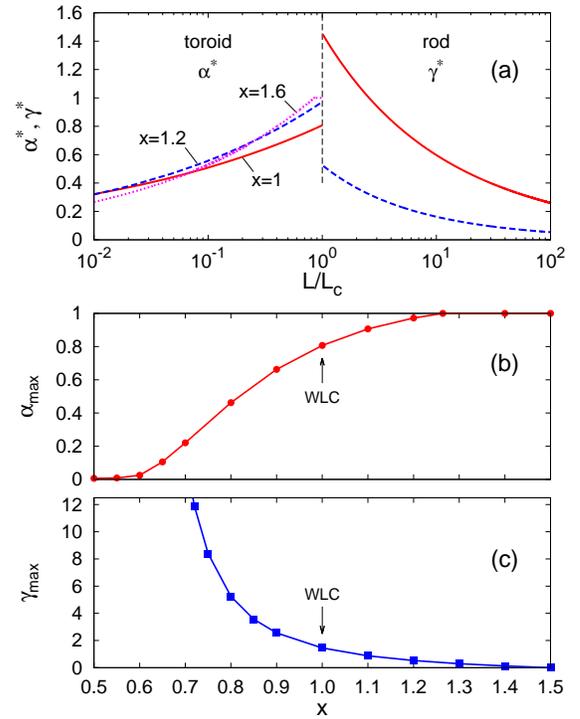}
\caption{\label{fig:toro}
(a) Dependence of geometrical parameters $\alpha$ and $\gamma$ on the chain
length $L$ for toroidal (left) and rod-like (right) condensates. At $L=L_c$ the
chain undergoes a transition from toroidal to rod-like conformation. The data
shown are obtained for $x=1$ (solid), $x=1.2$ (dashed) and $x=1.6$ (dotted) as
indicated.  The dependence of the maximum values $\alpha_\mathrm{max}$ and
$\gamma_\mathrm{max}$ of the parameters $\alpha$ and $\gamma$ on the exponent
$x$ are shown in (b) and (c), respectively.  The arrow indicates the result for
the WLC model ($x=1$), for which $\alpha_\mathrm{max} \approx 0.808$ and
$\gamma_\mathrm{max} \approx 1.45$.  $\alpha_\mathrm{max}$ increases from zero
at $x=0.5$ and reaches unity at $x\approx 1.264$, whereas $\gamma_\mathrm{max}$
becomes zero only at $x=1.5$.
}
\end{figure}

\begin{figure}
\center
\includegraphics[width=3.0in]{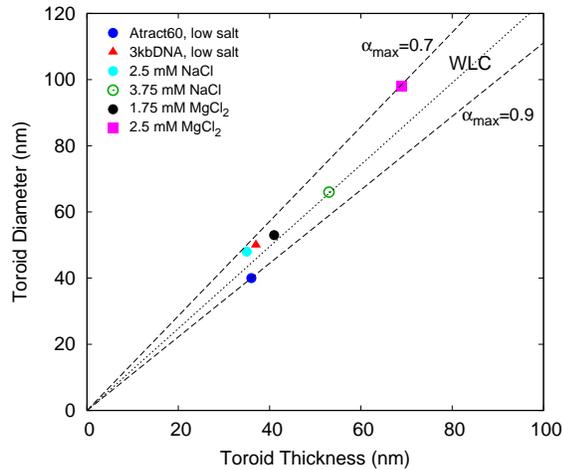}
\caption{\label{fig:hud}
Toroid thickness vs. toroid diameter as obtained by experiments at various
conditions (data points). The data points were extracted from Fig.  2 of Ref.
\cite{Conwell}. For a given condensation condition (see legend), we take only
the point with the maximum thickness to diameter ratio.  The lines show the
slopes corresponding to the maximum aspect ratios $\alpha_\mathrm{max}$ as
predicted from our model on changing the parameter $x$. For the worm-like chain
(WLC), $\alpha_\mathrm{max}=0.808$ (dotted line).
}
\end{figure}

Fig. \ref{fig:toro}a shows the dependence of $\alpha^*$ and $\gamma^*$ on $L$ away from the
transition line. At $L=L_c$ both $\alpha^*$ and $\gamma^*$ have maximum values,
$\alpha_\mathrm{max}$ and $\gamma_\mathrm{max}$, respectively, that strongly
depend on $x$.  For a given stiffness on increase of  $L$ the toroid condensate
becomes fatter and attains its maximum thickness at the transition line.  For
$x>1.246$ the toroid then attains its maximum thickness with the ratio
$\alpha=1$ even before $L$ reaches $L_c$ -- the donut has no hole (see the
curve for $x=1.6$ in Fig. \ref{fig:toro}a).  As $L$ becomes larger than $L_c$ the rod-like
condensate becomes the ground state, and eventually its length to thickness
ratio $\gamma^*$ starts decreasing. As $L \rightarrow \infty$, $\gamma^*$
becomes zero and the rod-like condensate turns into a globule.

At the transition line, both $L$ and $\kappa$ can be renormalized and the
maximum values $\alpha_\mathrm{max}$ and $\gamma_\mathrm{max}$ become functions
of the parameter $x$ only. Fig. \ref{fig:toro}(b and c) shows the dependence of
$\alpha_\mathrm{max}$ and $\gamma_\mathrm{max}$ on $x$, for $x \in [0.5,1.5]$.
For $x=0.5$, $\alpha_\mathrm{max} = 0$, while $\gamma_\mathrm{max}$ diverges,
indicating that both the toroidal and the rod-like phase disappear in this
limit.  $\alpha_\mathrm{max}$ increases with $x$ to unity at $x\approx 1.246$,
while $\gamma_\mathrm{max}$ decreases to zero as $x$ approaches 1.5.  This
again confirms that for $x \ge 1.5$ the rod-like condensate becomes a globule
for all lengths.  For the WLC model ($x=1$), $\alpha_\mathrm{max} \approx
0.808$ and $\gamma_\mathrm{max} \approx 1.45$.

It has been suggested that deviations from WLC occur at length scales less than
100 nm \cite{Wiggins}.  In Fig. \ref{fig:hud}, we show some experimental data
for the maximum thickness to diameter ratio (data extracted from Ref.
\cite{Conwell}) under various solution conditions. Experimental points
correspond to $\alpha_\mathrm{max}$ varying between 0.7 and 0.9 corresponding
to $x$ in the range from 0.9 to 1.1. Thus, our analysis suggests that the
solution conditions may be captured through the exponent $x$ in the generalized
elastic potential of DNA inside the condensate, but the bending elasticity
remains essentially that of the WLC, i.e. $x \approx 1 \pm 0.1$.  This suggests
that DNA compaction does not involve high deformation kinking of the chain,
consistent with various experimental studies on the looping of DNA (see e.g.
Ref. \cite{loop}) which indicated that the worm-like chain model is a
reasonable model for the elastic nature of DNA at short lengths.

\begin{acknowledgments}
This research is funded by Vietnam National Foundation for Science and
Technology Development (NAFOSTED) under grant number 103.01-2013.16.
\end{acknowledgments}

\end{document}